\documentclass[prd,twocolumn,groupedaddress,showpacs,nofootinbib]{revtex4}
\usepackage{graphicx}
\usepackage{dcolumn}
\usepackage{amssymb}
\usepackage{mathrsfs}
\usepackage{amsmath}
\usepackage{epsfig}
\usepackage[dvips]{color}
\usepackage{hhline}

\begin{document}

\title{An $SO(10)\times SO(10)'$ model for common origin of neutrino masses, ordinary and dark matter-antimatter asymmetries}

\author{Pei-Hong Gu}
\email{peihong.gu@sjtu.edu.cn}

\affiliation{Department of Physics and Astronomy, Shanghai Jiao Tong University, 800 Dongchuan Road, Shanghai 200240, China}

\begin{abstract}

We propose an $SO(10)\times SO(10)'$ model to simultaneously realize a seesaw for Dirac neutrino masses and a leptogenesis for ordinary and dark matter-antimatter asymmetries. A $(16\times \overline{16}')_{H}^{}$ scalar crossing the $SO(10)$ and $SO(10)'$ sectors plays an essential role in this seesaw-leptogenesis scenario. As a result of lepton number conservation, the lightest dark nucleon as the dark matter particle should have a determined mass around $15\,\textrm{GeV}$ to explain the comparable fractions of ordinary and dark matter in the present universe. The $(16\times \overline{16}')_{H}^{}$ scalar also mediates a $U(1)^{}_{em}\times U(1)'^{}_{em}$ kinetic mixing after the ordinary and dark left-right symmetry breaking so that we can expect a dark nucleon scattering in direct detection experiments and/or a dark nucleon decay in indirect detection experiments. If a proper mirror symmetry is imposed, our Dirac seesaw will not require more unknown parameters than the canonical Majorana seesaw.

\end{abstract}

\pacs{98.80.Cq, 14.60.Pq, 95.35.+d, 12.60.Cn, 12.60.Fr}

\maketitle

\section{Introduction}

In the most popular grand unified theories, we can naturally obtain the extremely light Majorana neutrinos through the famous seesaw mechanism \cite{minkowski1977,mw1980,flhj1989}. The lepton-number-violating interactions for the Majorana neutrinos can also accommodate a leptogenesis \cite{fy1986,luty1992,mz1992,fps1995,hms2000,di2002,hs2004,hlnps2004,hrs2005,dnn2008,bd2009} mechanism to explain the cosmic matter-antimatter asymmetry. However, the Majorana nature of the neutrinos is just a theoretical assumption and has not been confirmed experimentally. Meanwhile, all of the other observed fermions are the Dirac particles rather than the Majorana particles. Therefore, it is worth exploring the possibility of the Dirac neutrinos \cite{rw1983,mmps1987,gu2012,ct2013,lz2013,cgn2014,dlrw1999} in the grand unification framework.

On the other hand, the dark and ordinary matter contribute comparable energy densities in the present universe \cite{ade2013}. This coincidence can be understood in a nature way if the dark matter relic density is a dark matter-antimatter asymmetry \cite{nussinov1985,bcf1990,kaplan1992,dgw1992,kuzmin1998,kl2005,as2005,
clt2005,gsz2009,dmst2010,bdfr2011,mcdonald2011,hmw2010,dk2011,
frv2011,hms2011,kllly2011,gsv2011,fss2011,myz2012,idc2011,bpsv2011,cr2011,
as2012,ms2011,dm2012,fnp2012,bju2013,ams2013,cl2013} and has a common origin with the ordinary matter-antimatter asymmetry. The mirror world based on the gauge groups $[SU(3)^{}_c\times SU(2)^{}_L\times U(1)^{}_{Y}]\times [SU(3)'^{}_c\times SU(2)'^{}_L\times U(1)'^{}_{Y}]$ is a very attractive asymmetric dark matter scenario \cite{ly1956,kop1966,pavsic1974,bk1982,glashow1986,flv1991,flv1992,abs1992,hodges1993,bdm1996,silagadze1997,cf1998,
mt1999,bcv2001,bb2001,bgg2001,iv2003,fv2003,berezhiani2004,foot2004,berezhiani2006,
bb2006-1,bb2006,acmy2009,acmnz2010,dlnt2011,chly2012,gu2012,gu2013,pv2013,foot2013,zjm2013,foot2014}. The mirror models can contain a tiny $ U(1)^{}_{Y}\times U(1)'^{}_{Y}$ kinetic mixing input by hand to open a window for dark matter direct detections.

In this paper we shall propose an $SO(10)\times SO(10)'$ model with a $(16\times \overline{16}')_{H}^{}$ scalar to simultaneously realize a seesaw for Dirac neutrino masses and a leptogenesis for ordinary and dark matter-antimatter asymmetries. After the ordinary and dark left-right symmetry breaking, the $(16\times \overline{16}')_{H}^{}$ scalar can acquire an induced vacuum expectation value. The ordinary right-handed neutrinos and the dark left-handed neutrinos then can form three heavy Dirac fermions to highly suppress the masses between the ordinary left-handed neutrinos and the dark right-handed neutrinos. Meanwhile, these heavy Dirac fermions can decay to generate a lepton asymmetry in the ordinary leptons and an opposite lepton asymmetry in the dark leptons. The $SU(2)^{}_L$ and $SU(2)'^{}_R$ sphaleron processes respectively can transfer such lepton asymmetries to an ordinary baryon asymmetry and a dark baryon asymmetry. With calculable lepton-to-baryon conversations in the ordinary and dark sectors, the lightest dark nucleon as the dark matter particle should have a predictive mass about $15\,\textrm{GeV}$ to explain the ordinary and dark matter in the present universe as the ordinary proton has a known mass about $1\,\textrm{GeV}$. Benefited from the $U(1)^{}_{em}\times U(1)'^{}_{em}$ kinetic mixing mediated by the $(16\times \overline{16}')_{H}^{}$ scalar, the dark proton as the dark matter particle can scatter off the ordinary nucleons at a testable level while the dark proton/neutron as the dark matter particle can decay to produce the ordinary positron-electron pairs with a distinct energy. Compared to the canonical Majorana seesaw, our Dirac seesaw will not require more unknown parameters if a proper mirror symmetry is imposed.

\section{Fields and symmetry breaking}

In the ordinary $SO(10)$ sector, we have the fermions and scalars including
\begin{eqnarray}
\!\!\!\!&&\begin{array}{l}q^{c}_L(\mathbf{3},\mathbf{2},\mathbf{1},-\frac{1}{3})\end{array}\!\oplus \! \begin{array}{l}q^{}_R(\mathbf{3},\mathbf{1},\mathbf{2},+\frac{1}{3})\end{array}
\!\oplus\! \begin{array}{l}l^{c}_L(\mathbf{1},\mathbf{2},\mathbf{1},+1)\end{array}\nonumber\\
[1mm]
\!\!\!\!&&~~\oplus\! \begin{array}{l}l^{}_R(\mathbf{1},\mathbf{1},\mathbf{2},-1)=
\mathbf{16}^{}_F\,,\end{array}\nonumber\\
[2mm]
\!\!\!\!&&\begin{array}{l}\chi^{\ast}_L(\mathbf{1},\mathbf{2},\mathbf{1},+1)\end{array}\!\oplus\! \begin{array}{l}\chi^{}_R(\mathbf{1},\mathbf{1},\mathbf{2},-1)\in\mathbf{16}^{}_H\,,\end{array}\nonumber\\
[2mm]
\!\!\!\!&&\begin{array}{l}\Delta^{\ast}_L(\mathbf{1},\mathbf{3},\mathbf{1},-2)\end{array}\!\oplus \! \begin{array}{l}\Delta^{}_R(\mathbf{1},\mathbf{1},\mathbf{3},+2)\end{array}
\!\oplus\!\begin{array}{l}\Omega^{\ast}_L(\mathbf{3},\mathbf{3},\mathbf{1},-\frac{2}{3})\end{array}\nonumber\\
[1mm]
\!\!\!\!&&~~\oplus\! \begin{array}{l}\Omega^{}_R(\mathbf{3},\mathbf{1},\mathbf{3},+\frac{2}{3})
\in\mathbf{126}^{}_H\,,\end{array}\nonumber\\
[2mm]
\!\!\!\!&&\begin{array}{l}\Phi(\mathbf{1},\mathbf{2},\mathbf{2},0)\in\mathbf{10}^{}_H~\textrm{and/or~others}\,,\end{array}
\end{eqnarray}
where the brackets following the fields describe the transformations under the $SU(3)^{}_c\times SU(2)^{}_L\times SU(2)^{}_R \times U(1)^{}_{B-L}$ gauge groups.
Accordingly, the fermions and scalars in the dark $SO(10)'$ sector contain
\begin{eqnarray}
\!\!\!\!&&\begin{array}{l}q'^{c}_R(\mathbf{3},\mathbf{2},\mathbf{1},-\frac{1}{3})\end{array}\!\oplus \! \begin{array}{l}q'^{}_L(\mathbf{3},\mathbf{1},\mathbf{2},+\frac{1}{3})\end{array}
\!\oplus\!\begin{array}{l} l'^{c}_R(\mathbf{1},\mathbf{2},\mathbf{1},+1)\end{array}\nonumber\\
[1mm]
\!\!\!\!&&~~\oplus\! \begin{array}{l}l'^{}_L(\mathbf{1},\mathbf{1},\mathbf{2},-1)=
\mathbf{16}'^{}_F\,,\end{array}\nonumber\\
[2mm]
\!\!\!\!&&\begin{array}{l}\chi'^{\ast}_R(\mathbf{1},\mathbf{2},\mathbf{1},+1)\end{array}\!\oplus \! \begin{array}{l}\chi'^{}_L(\mathbf{1},\mathbf{1},\mathbf{2},-1)\in\mathbf{16}'^{}_H\,,\end{array}\nonumber\\
[2mm]
\!\!\!\!&&\begin{array}{l}\Delta'^{\ast}_R(\mathbf{1},\mathbf{3},\mathbf{1},-2)\end{array}\!\oplus \! \begin{array}{l}\Delta'^{}_L(\mathbf{1},\mathbf{1},\mathbf{3},+2)\end{array}
\!\oplus\!\begin{array}{l}\Omega'^{\ast}_R(\mathbf{3},\mathbf{3},\mathbf{1},-\frac{2}{3})\end{array}\nonumber\\
[1mm]
\!\!\!\!&&~~\oplus\! \begin{array}{l}\Omega'^{}_L(\mathbf{3},\mathbf{1},\mathbf{3},+\frac{2}{3})
\in\mathbf{126}'^{}_H\,,\end{array}\nonumber\\
[2mm]
\!\!\!\!&&\begin{array}{l}\Phi'(\mathbf{1},\mathbf{2},\mathbf{2},0)\in\mathbf{10}'^{}_H~\textrm{and/or~others}\end{array}\,,
\end{eqnarray}
where the brackets give the $SU(3)'^{}_c\times SU(2)'^{}_R\times SU(2)'^{}_L \times U(1)'^{}_{B-L}$ quantum numbers. There is also a $(\mathbf{16}\times \overline{\mathbf{16}'})_{H}^{}$ scalar crossing the $SO(10)$ and $SO(10)'$ sectors,
\begin{eqnarray}
\!\!\!\!\!\!\!\!(\mathbf{16}\times \overline{\mathbf{16}'})_{H}^{}&=&\Sigma_{l^{}_Ll'^{}_R}^{}
(\mathbf{1},\mathbf{2},\mathbf{1},-1)(\mathbf{1},\mathbf{2},\mathbf{1},+1)\nonumber\\[1mm]
&&\oplus
\Sigma_{l^{}_Rl'^{}_L}^{}
(\mathbf{1},\mathbf{1},\mathbf{2},+1)(\mathbf{1},\mathbf{1},\mathbf{2},-1)\oplus...\,.
\end{eqnarray}

For simplicity, we shall not consider the details of the $SO(10)$ and $SO(10)'$ symmetry breaking. Instead, we shall demonstrate at the left-right level. The ordinary and dark left-right symmetries are expected to have the breaking patterns as below,
\begin{subequations}
\label{ssb}\begin{eqnarray}
&&SU(3)^{}_c\times SU(2)^{}_L\times SU(2)^{}_R\times U(1)^{}_{B-L}  \nonumber\\
[1mm]
&&\stackrel{\langle\chi^{}_{R}\rangle=\frac{1}{\sqrt{2}}(v^{}_R,~0)^T_{}}
{-\!\!\!-\!\!\!-\!\!\!-\!\!\!-\!\!\!-\!\!\!-\!\!\!-\!\!\!-\!\!\!-\!\!\!-\!\!\!-\!\!\!-\!\!\!\longrightarrow} SU(3)^{}_c\times SU(2)^{}_L\times U(1)^{}_Y \nonumber\\
[1mm]
&&\stackrel{\langle\Phi\rangle=\frac{1}{\sqrt{2}}\textrm{diag}\{v^{}_1,~v^{}_2\}}
{-\!\!\!-\!\!\!-\!\!\!-\!\!\!-\!\!\!-\!\!\!-\!\!\!-\!\!\!-\!\!\!-\!\!\!-\!\!\!-\!\!\!-\!\!\!\longrightarrow} SU(3)^{}_c\times  U(1)^{}_{em}\,,\\
[4mm]
&&SU(3)'^{}_c\times SU(2)'^{}_R\times SU(2)'^{}_L\times U(1)'^{}_{B-L}  \nonumber\\
[1mm]
&&\stackrel{\langle\chi'^{}_{L}\rangle=\frac{1}{\sqrt{2}}(v'^{}_L,~0)^T_{}}
{-\!\!\!-\!\!\!-\!\!\!-\!\!\!-\!\!\!-\!\!\!-\!\!\!-\!\!\!-\!\!\!-\!\!\!-\!\!\!-\!\!\!-\!\!\!\longrightarrow} SU(3)'^{}_c\times SU(2)'^{}_R\times U(1)'^{}_Y \nonumber\\
[1mm]
&&\stackrel{\langle\Phi'\rangle=\frac{1}{\sqrt{2}}\textrm{diag}\{v'^{}_1,~v'^{}_2\}}
{-\!\!\!-\!\!\!-\!\!\!-\!\!\!-\!\!\!-\!\!\!-\!\!\!-\!\!\!-\!\!\!-\!\!\!-\!\!\!-\!\!\!-\!\!\!\longrightarrow} SU(3)'^{}_c\times  U(1)'^{}_{em}\nonumber\\
[1mm]
&&\stackrel{\langle\Delta'^{}_R\rangle=\frac{1}{\sqrt{2}}{\tiny\left[\begin{array}{cc}0&0\\
v'^{}_{em}&0\end{array}\right]}}
{-\!\!\!-\!\!\!-\!\!\!-\!\!\!-\!\!\!-\!\!\!-\!\!\!-\!\!\!-\!\!\!-\!\!\!-\!\!\!-\!\!\!-\!\!\!\longrightarrow} SU(3)'^{}_c\,.
\end{eqnarray}\end{subequations}
We further impose a $U(1)^{}_G$ global symmetry under which $(\chi^{\ast}_L,\chi^{}_R)$ and $(\chi'^{\ast}_R,\chi'^{}_L)$ carry a same charge. This means
the following cubic terms
\begin{eqnarray}
V\!\!&\supset& \!\!\rho^{}_{\Phi}\chi^\dagger_L\Phi\chi^{}_R +\widetilde{\rho}^{}_{\Phi} \chi^\dagger_L\widetilde{\Phi}\chi^{}_R
+\rho^{}_{\Phi'}\chi'^\dagger_R\Phi'\chi'^{}_L +\widetilde{\rho}^{}_{\Phi'} \chi'^\dagger_R\widetilde{\Phi'}\chi'^{}_L\nonumber\\
[1mm]
&&+\rho^{}_{\Delta}(\chi^{T}_Li\tau_2^{}\Delta^{}_L\chi^{}_L+\chi^{\dagger}_Ri\tau_2^{}\Delta^{\ast}_R\chi^{\ast}_R)
\nonumber\\
[1mm]
&&+\rho^{}_{\Delta'}(\chi'^{T}_Ri\tau_2^{}\Delta'^{}_R\chi'^{}_R+\chi'^{\dagger}_Li\tau_2^{}\Delta'^{\ast}_L\chi'^{\ast}_L)+\textrm{H.c.}\,,
\end{eqnarray}
should be absent from the scalar potential. Therefore the neutral components of the scalars $\chi^{}_L$, $\Delta^{}_{L,R}$, $\chi'^{}_R$, $\Delta'^{}_{R,L}$ will not acquire any induced vacuum expectation values. Accordingly, we can give a nonzero $\langle\Sigma^{}_{RL'}\rangle\leq\langle\chi^{}_{R,L'}\rangle$ and a zero $\langle\Sigma^{}_{LR'}\rangle$ from the scalar interactions as below,
\begin{eqnarray}
V&\supset& \rho^{}_{\Sigma}(\chi^T_L\Sigma^{\ast}_{LR'}\chi'^{\ast}_R + \chi^\dagger_R\Sigma^{}_{RL'}\chi'^{}_L)+\textrm{H.c.}\,.
\end{eqnarray}

\section{Dirac neutrinos and lepton asymmetries}

We write down the Yukawa couplings relevant for the fermion mass generation,
\begin{eqnarray}
\!\!\!\!\!\!\mathcal{L}\!\!&\supset&\!\! -y^{}_{q}\bar{q}^{}_L\Phi q^{}_R \!-\!\tilde{y}^{}_{q}\bar{q}^{}_L\widetilde{\Phi} q^{}_R \!-\!y^{}_{l}\bar{l}^{}_L\Phi l^{}_R \!-\!\tilde{y}^{}_{l}\bar{l}^{}_L\widetilde{\Phi} l^{}_R\nonumber\\
[1mm]
\!\!&&\!\! -y^{}_{q'}\bar{q}'^{}_R\Phi' q'^{}_L \!-\!\tilde{y}^{}_{q'}\bar{q}'^{}_R\widetilde{\Phi'} q'^{}_L \!-\!y^{}_{l'}\bar{l}'^{}_R\Phi' l'^{}_L \!-\!\tilde{y}^{}_{l'}\bar{l}'^{}_R\widetilde{\Phi'} l'^{}_L\nonumber\\
[1mm]
\!\!&&\!\!-\frac{1}{2}f^{}_{\Delta}(\bar{l}^{}_Li\tau_2^{}\Delta^{\ast}_{L}l^{c}_L\!+\!\bar{l}^{c}_Ri\tau_2^{}\Delta^{}_{R}l^{}_R)
\!-\frac{1}{2}f^{}_{\Delta'}(\bar{l}'^{}_Ri\tau_2^{}\Delta'^{\ast}_{R}l'^{c}_R\nonumber\\
[1mm]
\!\!&&\!\!+\bar{l}'^{c}_Li\tau_2^{}\Delta'^{}_{L}l'^{}_L)
\!-\!f^{}_\Sigma(\bar{l}^{c}_L\Sigma^{}_{l^{}_Ll'^{}_R}l'^{c}_R\!+\!\bar{l}^{}_R\Sigma^{}_{l^{}_Rl'^{}_L}l'^{}_L)
\!+\!\textrm{H.c.}\,.
\end{eqnarray}
When the left-right symmetries are broken down to the electroweak symmetries, we can derive
\begin{eqnarray}
\!\!\!\!\!\!\mathcal{L}\!\!&\supset&\!\! -y^{}_{u}\bar{q}^{}_L\phi u^{}_R \!-\!y^{}_{d}\bar{q}^{}_L\widetilde{\phi} d^{}_R \!-\!y^{}_{\nu}\bar{l}^{}_L\phi \nu^{}_R \!-\!y^{}_{e}\bar{l}^{}_L\widetilde{\phi} e^{}_R  \nonumber\\
[1mm]
\!\!&&\!\!-y^{}_{u'}\bar{q}'^{}_R\phi' u'^{}_L\!-\!y^{}_{d'}\bar{q}'^{}_R\widetilde{\phi'} d'^{}_L \!-\!y^{}_{\nu'}\bar{l}'^{}_R\phi' \nu'^{}_L \!-\!y^{}_{e'}\bar{l}'^{}_R\widetilde{\phi'} e'^{}_L\nonumber\\
[1mm]
\!\!&&\!\!-\frac{1}{2}f^{}_{\Delta}\bar{l}^{}_Li\tau_2^{}\Delta^{\ast}_{L}l^{c}_L
\!-\frac{1}{2}f^{}_{\Delta'}\bar{l}'^{}_Ri\tau_2^{}\Delta'^{\ast}_{R}l'^{c}_R\!-\!M^{}_{N}\bar{\nu}^{}_R \nu'^{}_L
\!+\!\textrm{H.c.}\nonumber\\
[2mm]
&&
\textrm{with}~\left\{\begin{array}{l}
y^{}_u=\frac{v^{}_1 y^{}_q +v^{}_2\tilde{ y}^{}_q}{\sqrt{v^2_1+v^2_2}}\,,~
y^{}_d=\frac{v^{}_2 y^{}_q +v^{}_1\tilde{ y}^{}_q}{\sqrt{v^2_1+v^2_2}}\,,\\
[4mm]
y^{}_{u'}=\frac{v'^{}_1 y^{}_{q'} +v'^{}_2\tilde{ y}^{}_{q'}}{\sqrt{v'^2_1+v'^2_2}}\,,~y^{}_{d'}=\frac{v'^{}_2 y^{}_{q'} +v'^{}_1\tilde{ y}^{}_{q'}}{\sqrt{v'^2_1+v'^2_2}}\,,\\
[4mm]
M^{}_{N}=\frac{1}{\sqrt{2}}f^{}_\Sigma v'^{}_L\,.
\end{array}\right.
\end{eqnarray}
Here the Higgs scalars $\phi$ and $\phi'$ with the vacuum expectation values,
\begin{eqnarray}
\langle\phi\rangle&=&\left[\begin{array}{c}\frac{1}{\sqrt{2}}v\\
[2mm]0\end{array}\right] \quad \left(v=\sqrt{v^2_1+v^2_2}\simeq 246\,\textrm{GeV}\right)\,,\nonumber\\
[2mm]
\langle\phi'\rangle&=&\left[\begin{array}{c}\frac{1}{\sqrt{2}}v'\\
[2mm]0\end{array}\right] \quad \left(v'=\sqrt{v'^2_1+v'^2_2}\right)\,,
\end{eqnarray}
are responsible for spontaneously breaking the ordinary and dark electroweak symmetries.

\begin{figure}
\vspace{4.0cm} \epsfig{file=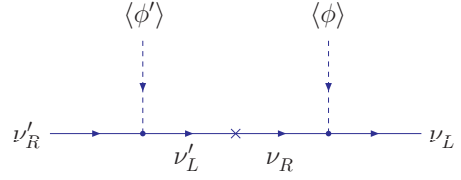, bbllx=5cm, bblly=6.0cm,
bburx=15cm, bbury=16cm, width=7cm, height=7cm, angle=0, clip=0}
\vspace{-8.25cm} \caption{\label{DiracI} The heavy masses between the ordinary right-handed neutrinos $\nu^{}_R$ and the dark left-handed neutrinos $\nu'^{}_L$ are responsible for suppressing the masses between the ordinary left-handed neutrinos
$\nu_L^{}$ and the dark right-handed neutrinos $\nu'^{}_R$.}
\end{figure}

According to the symmetry breaking pattern (\ref{ssb}), the fermion masses thus should be
\begin{eqnarray}
\!\!\!\!\mathcal{L}\!\!&\supset&\!\! - m^{}_{u}\bar{u}^{}_L u^{}_R - m^{}_{d}\bar{d}^{}_L d^{}_R - m^{}_{e}\bar{e}^{}_L e^{}_R -  m^{}_{u'}\bar{u}'^{}_R u'^{}_L\nonumber\\
[1mm]
\!\!&&\!\!- m^{}_{d'}\bar{d}'^{}_R d'^{}_L - m^{}_{e'}\bar{e}'^{}_R e'^{}_L -\frac{1}{2}\bar{m}^{}_{e'}\bar{e}'^{}_Re'^{c}_R\nonumber\\
[2mm]
\!\!&&\!\!
-\left[\begin{array}{cc}\bar{\nu}^{}_L&\bar{\nu}'^{}_L\end{array}\right]
\left[\begin{array}{cc}0&m_{LR}^{}\\
[2mm]
m_{R'L'}^{\dagger}&M_{N}^{\dagger}\end{array}\right]
\left[\begin{array}{c}\nu'^{}_R\\
[2mm]
\nu^{}_R\end{array}\right]+\textrm{H.c.}~~\textrm{with}\nonumber\\
[2mm]
\!\!&&\!\!m^{}_{f}\!=\!\frac{1}{\sqrt{2}}y^{}_{f}v\,,~
m^{}_{f'}\!=\!\frac{1}{\sqrt{2}}y^{}_{f'}v'\,,~
\bar{m}^{}_{e'}\!=\!\frac{1}{\sqrt{2}}f^{}_{\Delta'}v'^{}_{em}\,,\nonumber\\
[2mm]
\!\!&&\!\!m_{LR}\!=\!\frac{1}{\sqrt{2}}y^{}_{\nu}v\,,
~m_{R'L'}\!=\!\frac{1}{\sqrt{2}}y^{}_{\nu'}v'\,.
\end{eqnarray}
Note the dark charged leptons should be the so-called pseudo-Dirac particles for $v'^{}_{em}\ll v'$. As for the ordinary and dark neutrinos, their mass matrix can be block-diagonalized if the off-diagonal blocks are much lighter than the diagonal block,
\begin{eqnarray}
\mathcal{L}&\supset&-m_{\nu}^{}\bar{\nu}^{}_L\nu'^{}_R-M_N^{}\bar{\nu}^{}_R\nu'^{}_L+\textrm{H.c.}
~\textrm{with}\nonumber\\
[2mm]
&&\begin{array}{c}m_\nu^{}=-m_{LR}^{}\frac{1}{M^\dagger_N}m_{R'L'}^\dagger\,.\end{array}
\end{eqnarray}
Clearly, the ordinary left-handed neutrinos and the dark right-handed neutrinos can form the extremely light Dirac neutrinos as their masses are highly suppressed by the masses between the ordinary right-handed neutrinos and the dark left-handed neutrinos. This Dirac seesaw is definitely a variation of the canonical Majorana seesaw, see Fig. \ref{DiracI}. For the following discussions we can conveniently define the mass eigenstates by a proper phase rotation,
\begin{eqnarray}
\!\!N_i^{}\!=\!\nu^{}_{Ri}\!+\!\nu'^{}_{Li}~~\textrm{with}~~M_{N}^{}\!=\!\textrm{diag}\{M_{N_1^{}}^{},M_{N_2^{}}^{},M_{N_3^{}}^{}\}\,.
\end{eqnarray}

\begin{figure*}
\vspace{5.5cm} \epsfig{file=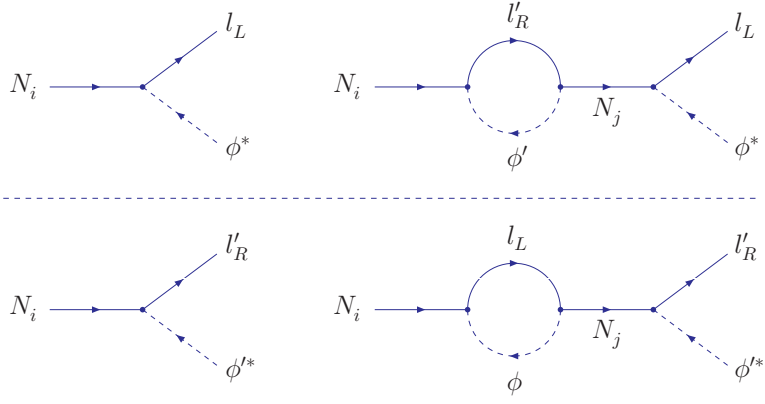, bbllx=5.5cm, bblly=6.0cm,
bburx=15.5cm, bbury=16cm, width=7cm, height=7cm, angle=0,
clip=0} \vspace{-7cm} \caption{\label{Ndecay} The lepton number
conserving decays of the heavy Dirac fermions
$N_i^{}=\nu^{}_{Ri}+\nu'^{}_{Li}$ into the ordinary leptons $l_L^{}$ as well as into the dark leptons $l'^{}_R$. The CP conjugation is not
shown for simplicity.}
\end{figure*}

As long as the CP is not conserved, the heavy Dirac fermions composed of the ordinary right-handed neutrinos and the dark left-handed neutrinos can have the lepton-number-conserving decays to generate a lepton asymmetry $\bar{\eta}^{}_L$ stored in the ordinary leptons and an opposite lepton asymmetry $\bar{\eta}'^{}_L$ stored in the dark leptons,
\begin{eqnarray}
\bar{\eta}^{}_L=-\bar{\eta}'^{}_L\propto \varepsilon_{N^{}_i}^{}\,.
\end{eqnarray}
Here $\varepsilon_{N^{}_i}^{}$ is the CP asymmetry defined as below,
\begin{eqnarray}
\varepsilon_{N^{}_i}^{}&=&\frac{\Gamma(N^{}_i\rightarrow l^{}_L\phi^\ast_{})-\Gamma(N^{}_i\rightarrow l^{c}_L\phi)}{\Gamma_{N^{}_i}^{}}\nonumber\\
&=&\frac{\Gamma(N^{}_i\rightarrow l'^{c}_R\phi')-\Gamma(N^{}_i\rightarrow l'^{}_R\phi'^\ast_{})}{\Gamma_{N^{}_i}^{}}~~~~\textrm{with}\nonumber\\
&&\begin{array}{ll}\Gamma_{N_i^{}}^{}&=\Gamma(N^{}_i\rightarrow l^{}_L\phi^\ast_{})+\Gamma(N^{}_i\rightarrow l'^{}_R\phi'^\ast_{})\\
[2mm]
&=\Gamma(N^{}_i\rightarrow l^{c}_L\phi)+\Gamma(N^{}_i\rightarrow l'^{c}_R\phi')\,.\end{array}
\end{eqnarray}
We can calculate the decay width at tree level,
\begin{eqnarray}
\Gamma_{N_i^{}}^{}=\frac{1}{16\pi}[(y^\dagger_{\nu}y^{}_\nu)_{ii}^{}+(y^\dagger_{\nu'}y^{}_{\nu'})_{ii}^{}]M_{N_i^{}}^{}\,,
\end{eqnarray}
and the CP asymmetry at one-loop level,
\begin{eqnarray}
\varepsilon_{N_i^{}}^{}=\frac{1}{4\pi}\sum_{j\neq i}\frac{\textrm{Im}[(y^{\dagger}_{\nu}y^{}_{\nu})_{ij}(y^{\dagger}_{\nu'}y^{}_{\nu'})_{ji}]}{(y^{\dagger}_{\nu}y^{}_{\nu})_{ii}
+(y^{\dagger}_{\nu'}y^{}_{\nu'})_{ii}}\frac{M_{N_i^{}}^{}M_{N_j^{}}^{}}{M_{N_i^{}}^2-M_{N_j^{}}^2}\,.
\end{eqnarray}
The relevant diagrams are shown in Fig. \ref{Ndecay}.

\section{Dark matter mass}

In the absence of other baryon asymmetries, the produced ordinary lepton asymmetry $\bar{\eta}^{}_{L}$ is equivalent to an ordinary $B-L$ asymmetry $\eta^{}_{B-L}=-\bar{\eta}^{}_{L}$ while the dark lepton asymmetry $\bar{\eta}'^{}_{L}$ is equivalent to a dark $B-L$ asymmetry $\eta'^{}_{B-L}=-\bar{\eta}'^{}_{L}$. The ordinary $SU(2)^{}_L$ sphaleron processes and the dark $SU(2)'^{}_R$ sphaleron processes then will partially transfer the ordinary and dark $B-L$ asymmetries to an ordinary baryon asymmetry $\eta^{}_B$ and a dark baryon asymmetry $\eta'^{}_B$, respectively,
\begin{subequations}
\begin{eqnarray}
\eta^{}_B&=&C\eta^{}_{B-L}=-C\bar{\eta}^{}_{L}~~\textrm{with}~~C=\frac{28}{79}\,,\\
\eta'^{}_B&=&C\eta'^{}_{B-L}=-C'\bar{\eta}'^{}_{L}~~\textrm{with}~~C'=\frac{28}{229}\,.
\end{eqnarray}
\end{subequations}
Note when computing the dark lepton-to-baryon conversation factor $C'$ we should take the $[SU(2)'^{}_{R}]$-triplet scalar $\Delta'^{}_R$ into account since this scalar drives the dark electromagnetic symmetry breaking much below the dark electroweak scale.

After the dark electromagnetic symmetry breaking, the dark charged leptons acquire a lepton-number-violating Majorana mass term so that the final dark charged lepton asymmetry cannot survive at all \cite{bp2011}. The lightest dark charged lepton denoted as the dark electron will leave a thermally produced relic density,
\begin{eqnarray}
\Omega^{}_{e'}h^2_{}\simeq \frac{0.1}{\langle\sigma^{}_{e'^{+}_{}e'^{-}_{}}v_{\textrm{vel}}^{}\rangle}~~\textrm{with}~~
\langle\sigma^{}_{e'^{+}_{}e'^{-}_{}}v_{\textrm{vel}}^{}\rangle =\frac{\pi\alpha'}{m^2_{e'}}\,.
\end{eqnarray}
Here $\alpha'$ is the dark fine-structure constant. It is easy to check the dark electron will only give a negligible relic density if its mass is at the GeV scale. Furthermore, we will show later the dark photon can efficiently decay into the ordinary fermion pairs. Therefore, if the lightest dark nucleon $N'$ is expected to serve as the dark matter particle, its mass should be determined by
\begin{eqnarray}
\!\!\!\!\!\!\!\!\!\!\!\!\!\!\!\!&&\Omega^{}_{\textrm{B}}h^2_{}:\Omega^{}_{\textrm{DM}}h^2_{}= \eta^{}_B m^{}_p : (-\eta'^{}_B) m^{}_{N'}\Rightarrow\nonumber\\
\!\!\!\!\!\!\!\!\!\!\!\!\!\!\!\!&& m_{N'}^{}\!=\!\frac{C'}{C}\frac{\Omega^{}_{\textrm{DM}}h^2_{}}{\Omega^{}_{\textrm{B}}h^2_{}}m^{}_p
\!=\!14.79\,\textrm{GeV}\!\left(\!\frac{\Omega^{}_{\textrm{DM}}h^2_{}/0.1199}{\Omega^{}_{\textrm{B}}h^2_{}/0.02205}\!\right)\,.
\end{eqnarray}

\section{Dark matter detection}

We can calculate the $U(1)^{}_{em}\times U(1)'^{}_{em}$ kinetic mixing at one-loop level,
\begin{eqnarray}
\mathcal{L}&\supset&-\frac{\epsilon}{2}A'^{}_{\mu\nu}A^{\mu\nu}_{}~~\textrm{with}\nonumber\\
&&\epsilon=\frac{\sqrt{\alpha\alpha'}}{12\pi}\sum_{Q,Q'}^{}QQ'C_{Q}^{}C_{Q'}^{}\ln\left[\frac{M^2_{(Q,Q')}}{\mu^2_{}}\right]\,.
\end{eqnarray}
Here $Q,Q'=\pm1,\pm\frac{1}{3},\pm\frac{2}{3}$ are the ordinary and dark electric charges of the scalars $\sigma(Q,Q')\in (\mathbf{16}\times \overline{\mathbf{16}'})_{H}^{}$, $M^{}_{(Q,Q')}$ denotes the $\sigma(Q,Q')$'s mass, $\mu$ is a renormalizable scale, while $C_{Q,Q'}^{}=1$ for $Q,Q'=\pm1$ and $C_{Q,Q'}^{}=3$ for $Q,Q'=\pm\frac{1}{3},\pm\frac{2}{3}$ are the color factors. Clearly, we have $\epsilon=0$ at the GUT scale. However, such kinetic mixing can appear after the left-right symmetry breaking,
\begin{eqnarray}
\!\!\!\!\!\!\!\!\epsilon\!\!&=&\!\! \frac{\sqrt{\alpha\alpha'}}{12\pi}\left[\ln\frac
{\left(1+\frac{1}{2}\lambda v'^2_L/M^2_1\right)\left(1+\frac{1}{2}\lambda v^2_R/M^2_1\right)}{1+\frac{1}{2}\lambda(v'^2_L+v^2_R)/M^2_1}\right.\nonumber\\
[1mm]
\!\!\!\!\!\!\!\!\!\!&&\!\!\left.+\ln\frac
{\left(1+\frac{1}{2}\lambda v'^2_L/M^2_3\right)\left(1+\frac{1}{2}\lambda v^2_R/M^2_3\right)}{1+\frac{1}{2}\lambda(v'^2_L+v^2_R)/M^2_3}\right]\nonumber\\
[1mm]
\!\!\!\!\!\!\!\!\!\!&\simeq&\!\! \frac{\sqrt{\alpha\alpha'}}{48\pi}\frac{\lambda^2_{}v'^2_Lv^2_R}{M^4_1}~~\textrm{for}~~M^2_3\gg M^2_1\gg\lambda v'^2_{L}\,,\lambda v^2_{R}\nonumber\\
[1mm]
\!\!\!\!\!\!\!\!\!\!&\sim&\!\! \frac{\sqrt{\alpha\alpha'}}{48\pi}\lambda^2_{} \!
=\! 10^{-9}_{}\!\left(\!\frac{\lambda}{0.0046}\!\right)^{\!2}_{}\!\!\sqrt{\frac{\alpha'}{\alpha}}~\textrm{for}~M^2_1\!\sim \!v'^{2}_L\!\sim\! v^2_R\,.\nonumber\\
&&
\end{eqnarray}
In the above calculation we have simplified the left-right level interactions as
\begin{eqnarray}
V&\supset& \lambda\left(\chi^\dagger_R\widetilde{\Sigma^{}}_{f^{}_Rf'^{}_L} \widetilde{\Sigma}^{\dagger}_{f^{}_Rf'^{}_L}\chi^{}_R
+\chi'^\dagger_L\widetilde{\Sigma}^{\dagger}_{f^{}_Rf'^{}_L} \widetilde{\Sigma}^{}_{f^{}_Rf'^{}_L}\chi'^{}_L\right.\nonumber\\
[1mm]
&&\left.+\chi^\dagger_R\Sigma^{}_{f^{}_Rf'^{}_L} \Sigma^{\dagger}_{f^{}_Rf'^{}_L}\chi^{}_R
+\chi'^\dagger_L\Sigma^{\dagger}_{f^{}_Rf'^{}_L} \Sigma^{}_{f^{}_Rf'^{}_L}\chi'^{}_L\right)\nonumber\\
[1mm]
&&+M^{2}_{1}\textrm{Tr}\left(\Sigma^\dagger_{l^{}_{R,L}l'^{}_{L,R}}\Sigma^{}_{l^{}_{R,L}l'^{}_{L,R}}\right)\nonumber\\
[1mm]
&&+M^{2}_{3}\sum_{f^{}_{R,L}\neq l^{}_{R,L}}^{f'^{}_{L,R}\neq l'^{}_{L,R}}\textrm{Tr}\left(\Sigma^\dagger_{f^{}_{R,L}f'^{}_{L,R}}\Sigma^{}_{f^{}_{R,L}f'^{}_{L,R}}\right)\,.
\end{eqnarray}

Due to the $U(1)^{}_{em}\times U(1)'^{}_{em}$ kinetic mixing, the physically dark photon will couple to not only the dark charged fermions but also the ordinary charged fermions although the physically ordinary photon doesn't couple to the dark charged fermions,
\begin{eqnarray}
\mathcal{L}&\supset&(\hat{A}^{}_\mu-\frac{\epsilon}{\sqrt{1-\epsilon^2_{}}} \hat{A}'^{}_\mu)(-\bar{e}\gamma^\mu_{}e-\frac{1}{3}\bar{d}\gamma^\mu_{}d+\frac{2}{3}\bar{u}\gamma^\mu_{}u)\nonumber\\
&&
+\hat{A}'^{}_\mu(-\bar{e}'\gamma^\mu_{}e'-\frac{1}{3}\bar{d}'\gamma^\mu_{}d'+\frac{2}{3}\bar{u}'\gamma^\mu_{}u')\,,
\end{eqnarray}
where the physical photons have been defined by \cite{fh1991}
\begin{eqnarray}
\hat{A}^{}_\mu=A^{}_\mu+\epsilon A'^{}_\mu\,,~~\hat{A}'^{}_\mu=\sqrt{1-\epsilon^2_{}}A'^{}_\mu\,.
\end{eqnarray}
Once the kinematics is allowed, the dark photon $\gamma'$ can efficiently decay into the ordinary charged fermion pairs,
\begin{eqnarray}
\Gamma^{}_{\gamma'\rightarrow f\bar{f}}=\frac{\epsilon^2_{}Q_f^2C^{}_Q}{12\pi}m_{A'}^{}\left(1-\frac{m_f^2}{m_{A'}^2}\right)
\sqrt{1-4\frac{m_f^2}{m_{A'}^2}}\,,
\end{eqnarray}
with the dark photon mass $m^{2}_{A'}=16\pi\alpha'v'^2_{em}$ and the ordinary electric charges $Q_{e,\mu,\tau}^{}=-1$, $Q_{d,s,b}^{}=-\frac{1}{3}$ and $Q_{u,c,t}^{}=+\frac{2}{3}$.

The dark photon can mediate an elastic scattering of the dark nucleons off the ordinary nucleons. If the dark proton is the dark matter particle, its scattering will have a spin-independent cross section,
\begin{eqnarray}\!\!\!\!\!\!\!\!\!\!\!\!&& \sigma_{p'N\rightarrow
p'N}^{}(Z,A)\nonumber\\
[1mm]
\!\!\!\!\!\!\!\!\!\!\!\!&\simeq &\epsilon^2_{}\pi
\alpha\alpha'\frac{[m_{p'}^{}m_{p}^{}/(m_{p'}^{}+m_{p}^{})]^2_{}}{m_{A'}^4}
\left(\frac{Z}{A}\right)^2_{}\nonumber\\
[2mm]\!\!\!\!\!\!\!\!\!\!\!\!& \simeq& 5.1\times10^{-46}_{}\,\textrm{cm}^2_{}\left(\frac{\alpha'}{\alpha}\right)\left(\frac{Z}{A}\right)^2_{}\nonumber\\
&&\times \left(\frac{\epsilon}{10^{-9}_{}}\right)^2_{}
\left(\frac{100\,\textrm{MeV}}{m_{A'}^{}}\right)^4_{}\,.
\end{eqnarray}
Such dark matter scattering can be verified in the direct detection experiments \cite{akerib2013}. If the dark neutron is the dark matter particle, its scattering off the ordinary nucleons will be far away from the experimental sensitivities \cite{acmnz2010}. In the present $SO(10)\times SO(10)'$ framework, we can expect a dark nucleon decay according to the ordinary proton decay. It should be noted the dark leptoquark scalars $\Omega'^{}_{R,L}$ can be allowed much lighter than the ordinary ones $\Omega_{L,R}^{}$. This means the dark nucleon decay can be fast enough to open a window for the indirect detection experiments although the ordinary proton decay is extremely slow. For example, in the dark matter decay chains $p'\rightarrow \pi'^0_{}e'^{+}_{}~(\textrm{or}~n'\rightarrow \pi'^0_{}\bar{\nu}'^{}_{R})$, $\pi'^0_{}\rightarrow \gamma'\gamma'$, $\gamma'\rightarrow e^{+}_{}e^{-}_{}$, the induced positrons/electrons can have a distinct energy,
\begin{eqnarray}
E_{e^{\pm}_{}}^{}&\simeq& \frac{m_{N'}^2+m_{\pi'^{0}_{}}^2}{8m_{N'}^{}}~~(\textrm{for}~~ m_{e'}^{}\ll m_{\pi'^{0}_{}}^{}<m_{N'}^{})\nonumber\\
&\simeq& 1.9\,\textrm{GeV}\left(\frac{m_{N'}^{}}{15\,\textrm{GeV}}\right)
\left[1+\left(\frac{m_{\pi'^{0}_{}}^{}}{m_{N'}^{}}\right)^2_{}\right]\nonumber\\
&\in&(1.9\textrm{GeV},~3.8\textrm{GeV})\,.
\end{eqnarray}
Clearly, if the dark photon mass is about $1-2\,\textrm{MeV}$, the dark matter should mostly decay into the positron/electron pairs.

\section{Discrete mirror symmetry}

We can impose a softly or spontaneously broken mirror symmetry under which the ordinary and dark fields transform as
\begin{eqnarray}
16^{}_{F}\longleftrightarrow16'^{c}_{F}\,,~~16^{}_{H}\longleftrightarrow16'^{}_{H}\,,...
\end{eqnarray}
to simplify the parameter choice,
\begin{eqnarray}
y^{}_{f}=y^{\ast}_{f'}\,,~~\tilde{y}^{}_{f}=\tilde{y}^{\ast}_{f'}\,,~~f^{}_{\Sigma}=f^{T}_{\Sigma}\,,...
\end{eqnarray}
By further assuming
\begin{eqnarray}
\frac{v'^{}_1}{v'^{}_2}=\frac{v^{}_1}{v^{}_2}\,,
\end{eqnarray}
we can read
\begin{eqnarray}
\label{dfmass}
\frac{\langle v'\rangle}{\langle  v\rangle}&=&\frac{m_{u'}^{}}{m_{u}^{}}
=\frac{m_{d'}^{}}{m_{d}^{}}=\frac{m_{s'}^{}}{m_{s}^{}}
=\frac{m_{c'}^{}}{m_{c}^{}}=\frac{m_{b'}^{}}{m_{b}^{}}
=\frac{m_{t'}^{}}{m_{t}^{}}\nonumber\\
&=&\frac{m_{e'}^{}}{m_{e}^{}}
=\frac{m_{\mu'}^{}}{m_{\mu}^{}}=\frac{m_{\tau'}^{}}{m_{\tau}^{}}\,.
\end{eqnarray}
We then can make use of the beta functions of the ordinary and dark QCDs to determine
\begin{eqnarray}
\label{dqcd}
\Lambda^{}_{\textrm{QCD}'}&=&\left(\frac{v'}{v}\right)^{\frac{4}{11}}_{}
(m_u^{}m_d^{}m_s^{})^{\frac{2}{33}}_{}\Lambda^{\frac{9}{11}}_{\textrm{QCD}}\nonumber\\
&&~~\textrm{for}~~\Lambda_{\textrm{QCD}'}^{}<m_{u'}^{}\,.
\end{eqnarray}
Since the dark hadronic scale is lighter than the dark quark masses, we can simply ignore the dark QCD contributions to the masses of the dark baryons and mesons such as
\begin{eqnarray}
\label{bmmass}
&&m_{p'}^{}\simeq 2m_{u'}^{}+m_{d'}^{}\,,~~m_{n'}^{}\simeq m_{u'}^{}+2m_{d'}^{}\,,\nonumber\\
&&m_{\pi'}^{}\simeq m_{\pi'^{0}_{}}^{}\simeq m_{\pi'^{\pm}_{}}^{}\simeq m_{u'}^{}+m_{d'}^{}\,.
\end{eqnarray}
From Eqs. (\ref{dfmass}-\ref{bmmass}), we can obtain
\begin{eqnarray}
&&m_{e'}^{}=1.5\,\textrm{GeV}\,,~~m_{u'}^{}=3.75\,\textrm{GeV}\,,~~m_{d'}^{}=7.5\,\textrm{GeV}\,,\nonumber\\
&&\Lambda_{\textrm{QCD}'}^{}=2\,\textrm{GeV}\,,~~m_{p'}^{}=15\,\textrm{GeV}\,,~~m_{n'}^{}=18.75\,\textrm{GeV}\,,\nonumber\\
&&m_{\pi'}^{}=11.25\,\textrm{GeV}\,,
\end{eqnarray}
by inputting
\begin{eqnarray}
\!\!\!\!\!\!\!\!\!\!\!\!\!\!\!\!&&v'=3000\,v\,,~m_{e}^{}=0.511\,\textrm{MeV}\,,~m_{u}^{}=1.25\,\textrm{MeV}\,,\nonumber\\
\!\!\!\!\!\!\!\!\!\!\!\!\!\!\!\!&&m_{d}^{}=2.5\,\textrm{MeV}\,,~m_{s}^{}=100\,\textrm{MeV}\,,~\Lambda_{\textrm{QCD}}^{}=200\,\textrm{MeV}\,.
\end{eqnarray}
In this case, the dark proton is the lightest dark nucleon and hence is the dark matter particle. Another interesting consequence of this mirror symmetry is that our Dirac seesaw doesn't contain more unknown parameters than the canonical Majorana seesaw.

\section{Summary}

In this paper we have proposed an $SO(10)\times SO(10)'$ model to simultaneously explain the smallness of the Dirac neutrino masses and the coincidence between the ordinary and dark matter. Specifically we introduced a $(16\times \overline{16}')_{H}^{}$ scalar crossing the ordinary $SO(10)$ sector and the dark $SO(10)'$ sector. This $(16\times \overline{16}')_{H}^{}$ scalar can acquire an induced vacuum expectation value after the $16^{}_{H}$ and $16'^{}_{H}$ scalars drive the spontaneous breaking of the ordinary and dark left-right symmetries. Consequently the ordinary right-handed neutrinos and the dark left-handed neutrinos can form the heavy Dirac fermions to highly suppress the masses between the ordinary left-handed neutrinos and the dark right-handed neutrinos. The decays of such heavy Dirac fermions can generate an ordinary lepton asymmetry and an opposite dark lepton asymmetry. We hence can obtain an ordinary baryon asymmetry and a dark baryon asymmetry due to the $SU(2)^{}_L$ and $SU(2)'^{}_R$ sphaleron processes.  By taking into account the difference between the ordinary and dark lepton-to-baryon conversations, we can expect the lightest dark nucleon as the dark matter particle to have a determined mass around $15\,\textrm{GeV}$. Furthermore, the $(16\times \overline{16}')_{H}^{}$ scalar can mediate a small $U(1)^{}_{em}\times U(1)'^{}_{em}$ kinetic mixing after the ordinary and dark left-right symmetry breaking. Therefore, the dark proton as the dark matter particle can be verified by the direct and indirect detection experiments. Alternatively, if the dark neutron is the dark matter particle, it can be only found by the indirect detection experiments. In particular, the dark nucleon decay can produce the ordinary positron-electron pairs with a distinct energy about $1.9-3.8\,\textrm{GeV}$. Compared to the canonical Majorana seesaw, our Dirac seesaw will not require more unknown parameters if a proper mirror symmetry is imposed.

\textbf{Acknowledgement}: This work was supported by the Shanghai Jiao Tong University under Grant No. WF220407201 and the Shanghai Laboratory for Particle Physics and Cosmology under Grant No. 11DZ2260700.

\end{document}